\documentclass{aa}  
\usepackage{graphicx}
\usepackage{txfonts}
\usepackage{color}

\begin{document} 

\renewcommand {\arraystretch}{1.4}

   \title{Quasi-periodic oscillations in accreting magnetic white dwarfs}
   \subtitle{II. The asset of numerical modelling for interpreting observations}
    \titlerunning{Numerical modelling of QPOs in polars}
   \author{C. Busschaert \inst{1,2} 
   \and \'E. Falize \inst{2} \and C. Michaut \inst{1} \and J.-M. Bonnet-Bidaud \inst{3} \and M. Mouchet \inst{1}}
   \institute{
   LUTH, Observatoire de Paris, UMR CNRS 8102, Universit\'e Paris Diderot, 92190 Meudon, France
   \and
   CEA-DAM-DIF, F-91297 Arpajon, France
   \and
   CEA Saclay, DSM/Irfu/Service d'Astrophysique, 91191 Gif-sur-Yvette, France
   }
   \offprints{clotilde.busschaert@cea.fr}      
   \date{  \date{Received: 8 December 2014; Accepted: 26 May 2015} }

 
  \abstract
   {Magnetic cataclysmic variables are close binary systems containing a strongly magnetized white dwarf that accretes matter coming from an M-dwarf companion. The high magnetic field strength leads to the formation of an accretion column instead of an accretion disk. High-energy radiation coming from those objects is emitted from the column close to the white dwarf photosphere at the impact region. Its properties depend on the characteristics of the white dwarf and an accurate accretion column model allows the properties of the binary system to be inferred, such as the white dwarf mass, its magnetic field, and the accretion rate.}
   {We study the temporal and spectral behaviour of the accretion region and use the tools we developed to accurately connect the simulation results to the X-ray and optical astronomical observations.}
   {The radiation hydrodynamics code \textsc{Hades} was adapted to simulate this specific accretion phenomena. Classical approaches were used to model the radiative losses of the two main radiative processes: bremsstrahlung and cyclotron. Synthetic light curves and X-ray spectra were extracted from numerical simulations. A fast Fourier analysis was performed on the simulated light curves. The oscillation frequencies and amplitudes in the X-ray and optical domains are studied to compare those numerical results to observational ones. Different dimensional formulae were developed to complete the numerical evaluations. }
   {The complete characterization of the emitting region is described for the two main radiative regimes: when only the bremsstrahlung losses and when both cyclotron and bremsstrahlung losses are considered. The effect of the non-linear cooling instability regime on the accretion column behaviour is analysed. Variation in luminosity on short timescales ($\sim$ ~1~s quasi-periodic oscillations) is an expected consequence of this specific dynamic. The importance of secondary shock instability on the quasi-periodic oscillation phenomenon is discussed. The stabilization effect of the cyclotron process is confirmed by our numerical simulations, as well as the power distribution in the various modes of oscillation.}
{}

   \maketitle
%

\section{Introduction}

Magnetic cataclysmic variables are close binary systems containing a highly magnetized  (B $\sim 10-200$~MG) white dwarf that accretes matter from a low-mass companion. For a subclass of those objects, called AM Her stars or polars, the magnetic field of the white dwarf is strong enough to lock the system into a synchronous rotation and to prevent the formation of an accretion disk \citep{Cro90, War95}. Matter leaves the companion, follows a ballistic trajectory, and is then captured by the magnetic field lines of the white dwarf.  Its  falls  down to the magnetic pole of the compact star at supersonic free-fall velocity ($v_{\textrm{ff}} \sim 5-10\times10^8$~cm/s), forming an accretion column structure. \\

The collision of the infalling flow with the white dwarf photosphere leads to forming a reverse shock, which propagates upstream and heats the plasma to temperature $T_{\textrm{s}} \simeq 10-50$~keV. These temperatures are due to the high accretion velocity ($T_{\textrm{s}} \propto v_{\textrm{ff}}^2$), which is directly linked to the mass (and radius) of the white dwarf. Those extreme temperatures, as well as the strong magnetic field, imply that the post-shock region radiates through several radiative processes from the optical to X-ray domains. The radiation arising from these binary systems comes mainly from the accretion region. As a consequence, these systems can be used as a perfect probe for testing accretion models in high-energy regimes. The energy losses shape the density and temperature profiles in the post-shock region and induce a strong gradient of density, temperature, and velocity near the white dwarf surface. Since the radiation properties depend on the mass of the white dwarf, modelling the emitting region can be used to determine the physical properties of the compact object \citep[see][]{Kin79, Cro98}. Numerous studies have shown, in fact, that modelling of the post-shock region physics strongly affects any determination of the white dwarf properties \citep{Ish91, Wu95, Yu10, Hay14}. \\

A stationary regime is frequently assumed to be established to help in interpreting observations \citep{Fis01}. However, radiative instabilities that can potentially develop in the post-shock region can induce dynamical effects and lead to the unsteady behaviour of the  post-shock accretion column (hereafter PSAC). Indeed, the efficiency of the bremsstrahlung process can trigger the cooling instability. The shock thus oscillates around the stationary position.  First numerical studies of such accretion flows have highlighted the presence of the cooling instability in the column \citep{Lan81} supported by theoretical studies \citep{Che82}. A comprehensive review of the different stationary and instationary models was given later by \cite{Wu00}. \\

The oscillation of the front shock leads to a variation in the luminosity due to the induced variation of the emission volume, as well as of the conditions of density and temperature inside the column \citep{Ima91, Wu92}. The characteristic timescale of this phenomenon is the cooling timescale associated to the bremsstrahlung emission, i.e. $t_{\textrm{brem}} \sim 1$~s. This timescale value is consistent with the periods  of the quasi-periodical oscillations (QPOs) observed in the optical light curves of a few polars \citep{Mid82, Mas83, BB85, Lar87, Lar89, BB96}. These rapid oscillations are characterized by an amplitude of 1-5\% rms and a period of 1-2.5~s. From the model of cooling instability, an X-ray counterpart of optical oscillations is theoretically expected. Thus QPOs have been extensively searched in X-ray light curves. New results are presented in a related paper (\cite{BB14}, hereafter referred as Paper I).   \\

In this context, various analytical and numerical studies have been performed to determine the sensitivity of the instability to physical parameters, such as the reverse shock Mach number \citep{Str95, Ram06} or the influence of different possible physical processes taking part in the post-shock region, like two-temperature effects \citep{Ima96, Sax99, Sax05} or transverse magnetic field \citep{Tot93, Ram05}. A few studies have also described the effect of the coupling between several cooling processes \citep{Wu92, Sax98}. In particular, recent astronomical observations (see Paper I) show that several polars are in regimes where either the bremsstrahlung dominates the losses or both bremsstrahlung and cyclotron have to be taken into account. \\

The two main conclusions of those previous works are the following: Firstly, the cyclotron losses damp the oscillation. Thus, a critical magnetic field amplitude can be found for which the oscillation disappears after a few periods. To sustain an oscillation for a longer time, a noisy accretion rate has been invoked \citep{Wol91, Wood92, Wu92}. Secondly, the oscillations of the emission in the X-ray domain due to the bremsstrahlung process and in the optical domain due to the cyclotron one are expected to present the same frequencies and the same order of magnitude of the relative amplitudes \citep{Wu92}. To complement those results and to better understand the physics of this specific regime, we present new numerical simulations to study the unsteady behaviour of the accretion column. Those simulations are compared to observational results (see Paper I).  \\ 

In Section \ref{model}, the standard model is reviewed in order to build the parameter space and the relevant accretion regimes of the  column as a function of the three leading parameters (specific accretion rate, white dwarf mass, and magnetic field). In Paper I, four parameters are considered because the specific accretion rate is not reached directly by observations but through the combination of the total accretion rate and the column cross-section.  In Section \ref{non-stationnaire}, the general behaviour of a column in an unsteady state is studied with particular attention to the additional instabilities that can develop in the post-shock region. Then in Section \ref{parametre},  the dependency of the observational QPO parameters on the three leading parameters is developed further and discussed.   \\

\section{Models of radiation and magnetic processes in high-energy accretion regions}
\label{model}

\subsection{Brief review of the accretion column standard model}

In the standard model of accretion column, the matter coming from the companion is trapped by the gravitational field of the compact object and falls at free-fall velocity, as given by \mbox{$v_{\textrm{ff}}=(2 G M_{\textrm{WD}}/R_{\textrm{WD}})^{1/2}$},  onto the magnetic poles of the white dwarf, with $G$, $M_{\textrm{WD}}$, and $R_{\textrm{WD}}$ being the gravitational constant, the white dwarf mass, and radius, respectively. It should be noted that the matter is actually captured at the Alfven radius, $R_A$, so the free fall velocity should be corrected by a factor $(1-R_{\textrm{WD}}/R_{A})$. However, for polars, the Alfven radius is much larger than the white dwarf radius so this effect has been neglected.  Considering the Nauenberg mass-radius relation  \citep{Nau72}, the free-fall velocity only depends on the white dwarf mass. Typically, this velocity  is $v_{\textrm{ff}} \sim 8.75\times10^8$ cm/s for a 1 $M_{\odot}$ white dwarf. The incoming flow is thus highly supersonic with a Mach number of $\mathcal{M} \sim (10-100)$. \\

Owing to the magnetic collimation, matter spreads over a small fraction of the white dwarf surface, denoted $f$ and defined by $f=S/(4\pi R^{2}_{\textrm{WD}})$ where $S$ is the area of the column cross-section. This quantity is not well constrained by observations. As an example, this quantity has been evaluated in the range  $f \sim 10^{-5} - 3.10^{-4}$ by \citet{Ima84b} for a sample of polars. The impact of the accreting flow onto the white dwarf surface leads to the formation of a reverse shock, which counter-propagates in the incoming plasma. The shock transforms the kinetic energy into thermal energy, and the PSAC is strongly heated to high-temperature $T_{\textrm{s}}$. Using the Rankine-Hugoniot conditions and assuming that the accreted medium is described by the perfect gas approximation, the post-shock temperature scales as $v_{\textrm{ff}}^2$ and is given by 
\begin{equation}
T_{\textrm{s}} = \frac{\mu m_{\textrm{H}}}{k_{\textrm{B}}} \frac{2(\gamma-1)}{(\gamma+1)^2} v_{\textrm{ff}}^2 = 1.1 \times 10^7 \textrm{ K}\;\left( \frac{v_{\textrm{ff}}}{10^8 \textrm{ cm/s}} \right)^2, 
\end{equation}
with $\mu$, $m_{\textrm{H}}$, $k_{\textrm{B}}$, and $\gamma$ being the mean molecular weight, the hydrogen mass, the Boltzmann constant, and the adiabatic index, respectively. We use $\mu = 0.5$ (fully ionized hydrogen) throughout the paper and $\gamma=5/3$.\\

At such high temperatures, matter radiates strongly through optically thin bremsstrahlung, which dominates the hard X-ray emission \citep{Fra02}. Moreover, for the highest magnetized polars, the energy can also be transported efficiently by the cyclotron process \citep{Sax98}, which emits mainly in the optical domain for the range of polar magnetic fields. As infalling material cools and slows down, it also gets denser with a sharp gradient near the surface. Thanks to the energy losses by radiation, the reverse shock position eventually stabilizes. Consequently, when using the steady state approximation and a cooling dominated by bremsstrahlung losses, the height of post-shock accretion region can be estimated and is given by \cite{Wu94} as
\begin{equation}
x_{\textrm{s}} =  3 \times 10^7 \textrm{ cm}\ \left[ \frac{\dot{m}}{1 \textrm{ g/cm}^{2}\textrm{/s}}\right]^{-1} \left[ \frac{M_{\textrm{WD}}}{0.5 \textrm{ M}_{\odot}}\right]^{3/2}\left[\frac{R_{\textrm{WD}}}{10^9 \textrm{ cm}} \right]^{-3/2},\label{xs}
\end{equation}
with $\dot{m} = \rho_0 v_{\textrm{ff}}$ the specific accretion rate. A more complex but exact analytic expression can be found in \cite{Fal09}. With the Nauenberg relation, the relation (\ref{xs}) can be written as a function of $\dot{m}$ and $M_{\textrm{WD}}$. Given the order of magnitude of the accretion shock height, the local white dwarf gravitational field $g$ does not vary enough to strongly modify the post-shock structure ($g x_{\textrm{s}} << v_{\textrm{ff}}$). This approximation is valid as long as the white dwarf mass is below 1~M$_{\odot}$ \citep{Cro99}.\\
  
On the whole, three key parameters determine the post-shock structure: the white dwarf mass, M$_{\textrm{WD}}$, determining the free-fall velocity and the post-shock temperature; the specific accretion rate, $\dot{m}$, determining the post-shock density; and the local magnetic field strength, B, which determines the relative importance of cyclotron over bremsstrahlung losses. By comparing the characteristic timescales of the different physical processes at the front shock, the possible post-shock regimes are defined \citep{Lam79,Lan82}.\\
Four main regimes are obtained and can be explicitly defined in the (M$_{\textrm{WD}}$, $\dot{m}$, B) diagram. Figure 1  displays these regimes for a $0.8~\textrm{M}_{\odot}$ white dwarf in the ($\dot{m}$, B) plane. They are delimited by lines corresponding to an equal balance between two processes such as bremsstrahlung-cyclotron, electron-ion collisions, and ion-ion collisions. In the following, the characteristic timescales of these processes are recalled and used to derive the ($\dot{m}$, B) relations corresponding to the limits of the regimes.\\

Among all the radiative effects occurring in dense and hot magnetized plasmas, the dominant processes that shape the PSAC structure are the bremsstrahlung and cyclotron emissions \citep{Lam79,Wu00}. Through the estimate of optical depth of the post-shock medium to bremsstrahlung photons \citep{Fra02}, it can be deduced that radiation escapes the matter freely. Thus the loss of energy can be modelled by a cooling term that can be analytically expressed \citep{Ryb79} as  
\begin{equation}
\Lambda_{brem} = \frac{32/3}{(4\pi\varepsilon_{0})^{3}}\sqrt{\frac{2\pi^{3}k}{m_e^{3}}}\frac{Z^{2}e^{6}}{hc^{3}}n_{e}n_{i}g_{B}(Z,T)\sqrt{T}  \equiv \Lambda_0 \rho^{2} \textrm{T}^{1/2}\;, \label{brem}
\end{equation}
where $g_{B}$, $m_{e}$, $e$, $Z$, $n_e$, $n_i$, and $\rho$ are the Gaunt factor, the electron mass, the electron charge, the mean charge number, the electronic and ionic density, and the gas density, respectively. Assuming that the plasma consists of fully ionized hydrogen, the parameter $\Lambda_0$ in \eqref{brem} is estimated to  $\Lambda_0 \sim 5 \times 10^{20}$~erg.cm$^{3}$.g$^{-2}$.K$^{-1/2}$.s$^{-1}$. According to its definition in \cite{Ryb79}, the Gaunt factor is taken as $g_{B} \simeq 1$ (cf. \cite{Sax99}). \\

The bremsstrahlung cooling timescale, noted $t_{\textrm{brem}}$ is given by
\begin{eqnarray}
t_{\textrm{brem}}  & = &  \frac{\rho e}{ \Lambda_{\textrm{brem}}} \nonumber \\
t_{\textrm{brem}}  & \simeq &  4 \times 10^{-2} \textrm{ s}  \left( \frac{\dot{m}}{1 ~\textrm{g/cm}^2\textrm{/s}} \right)^{-1}  \left( \frac{v_{\textrm{ff}}}{10^8~\textrm{cm/s}} \right)^{2} , \label{t_brem} 
\end{eqnarray} 
with $e$ the internal energy per unit of mass. Parameters in relation \eqref{t_brem} have been normalized to the typical values encountered in polars. Thus, for a white dwarf of $0.8 $ M$_{\odot}$ with $\dot{m}\sim 1 ~\textrm{g/cm}^2\textrm{/s}$, the cooling timescale associated to the bremsstrahlung process is $t_{\textrm{brem}}\sim 1$ s. \\

When the local magnetic field close to the white dwarf surface is strong enough, the cyclotron process can also be an efficient radiative process for cooling the medium. Depending on the radiation frequency, the plasma is optically thick or optically thin to photons which strongly complexifies the modelling \citep{Lan82, Kal05}. Theoretical simplifications and dimensional considerations allow the cyclotronic transport to be reduced to a cooling function \citep{Lan82,Sax}, expressed as a function of the column cross-section area (S), the magnetic field, and the local thermodynamical conditions as 
\begin{eqnarray}
\Lambda_{\textrm{cycl}} & = &  1.2\times 10^{8} \textrm{ erg/cm}^3\textrm{/s } \left(\frac{S }{10^{15} \;\textrm{cm}^2}\right)^{-17/40}\left(\frac{B}{10 \;\textrm{MG}}\right)^{57/20} \nonumber\\
& &\times \left(\frac{\rho}{4. 10^{-8} \;\textrm{g/cm}^{3}}\right)^{3/20}\left(\frac{T}{10^8 \;\textrm{K}}\right)^{5/2} \;. \label{lambda_cyc}
\end{eqnarray}
For a given accretion column surface and a given white dwarf magnetic field, \eqref{lambda_cyc} is reduced to
\begin{eqnarray}
\Lambda_{\textrm{cycl}} & \equiv &  \Lambda_{0,c} \; \rho^{3/20}\;  T^{5/2}\;, \label{cyc}
\end{eqnarray}
where $\Lambda_{0,c}$ is thus a constant. \\

As for the bremsstrahlung, the cyclotron cooling timescale is given by
\begin{eqnarray}
t_{\textrm{cycl}}  &= & \frac{\rho e}{ \Lambda_{\textrm{cycl}}} \; \nonumber \\
t_{\textrm{cycl}}  & \simeq & 2 \times 10^2 \textrm{ s}\, \left( \frac{S}{10^{15}~\textrm{cm}^2} \right)^{17/40} \left( \frac{B}{10~\textrm{MG}} \right)^{-57/20} \nonumber \\
&& \times \left( \frac{\dot{m}}{1~\textrm{g/cm}^2\textrm{/s}} \right)^{17/20}  \left( \frac{v_{\textrm{ff}}}{10^8~\textrm{cm/s}} \right)^{-77/20}.\label{tcyc}
\end{eqnarray}
As expected, the cyclotronic effects are more efficient when the magnetic field is stronger and the plasma is hotter. The optically thick property of the medium can be retrieved from \eqref{tcyc}, since  $t_{\textrm{cycl}}$ increases with the surface and the density of the column. Indeed, the wider and denser  the column, the longer it takes the radiation to escape the column. \\
    
\begin{figure}[!ht]
\center
\includegraphics[width=\linewidth]{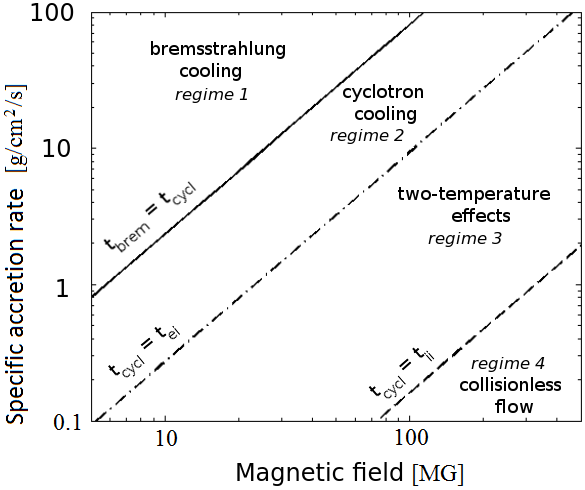}
\caption{Different regimes of accretion as a function of the accretion rate and the magnetic field strength, considering a white dwarf of $M=0.8$~M$_{\odot}$ and an accretion section of $S=10^{15}$~cm$^2$.}\label{busschaert_regime}
\end{figure}

To determine the prevailing mechanism between bremsstrahlung and cyclotron processes, the two cooling timescales computed right below the shock front are compared  through a dimensionless ratio, noted $\epsilon_{\textrm{s}}$ \citep{Wu94}, and given by
\begin{eqnarray}
\epsilon_{\textrm{s}} & \simeq  & \frac{t_{\textrm{brem}} }{t_{\textrm{cycl}}} =  2 \times 10^{-4} \left( \frac{S}{10^{15}~\textrm{cm}^2} \right)^{-17/40} \nonumber \\
& & \times \left( \frac{B}{10~\textrm{MG}} \right)^{57/20} \left( \frac{\dot{m}}{1~\textrm{g/cm}^2\textrm{/s}} \right)^{-37/20}  \left( \frac{v_{\textrm{ff}}}{10^8~\textrm{cm/s}} \right)^{117/20} \,. \label{epsilon_s}
\end{eqnarray}
The bremsstrahlung is the dominating process when \mbox{$\epsilon_{\textrm{s}}<1$} near the front shock, while the cyclotron is the dominant one when $\epsilon_{\textrm{s}}>1$. Since $\epsilon_{\textrm{s}}\propto B^{57/20}\dot{m}^{-37/20}$, the first regime is obtained for high accretion rates and weak magnetic fields (\emph{Regime 1} in Fig.\ref{busschaert_regime}), and the second regime corresponds to strong
 magnetic fields and low accretion rates (\emph{Regime 2} in Fig.\ref{busschaert_regime}). \\

From Eq. \eqref{epsilon_s} a critical magnetic field depending on $\dot{m}$, $S$, and $v_{\textrm{ff}}$ can be obtained at the limit $\epsilon_{\textrm{s}}=1,$ which separates the two regimes and is computed as 
\begin{eqnarray}
B_{\epsilon_{\textrm{s}}=1}  & \simeq & 200~ \textrm{ MG}\; \left(\frac{\dot{m}}{1\textrm{ g/cm}^2\textrm{/s}} \right)^{37/57} \left(\frac{v_{\textrm{ff}}}{10^8\textrm{ cm/s}} \right)^{-39/19} \nonumber \\ 
&& \qquad \times \left(\frac{S}{10^{15}\textrm{ cm}^2} \right)^{17/114}.
\end{eqnarray}
With the free-fall velocity expression and the white dwarf mass-radius relation, $B_{\epsilon_{\textrm{s}}=1}$ can be written as only a function of $\dot{m}$, $S$, and $M_{\textrm{WD}}$ . It should be emphasized that the bremsstrahlung losses increase towards the surface, while the cyclotron losses decrease (see section \ref{cyclotron_simulation}). In fact the radiative losses lead to the cooling of the plasma and its densification. This evolution implies an increase in the cyclotron cooling timescale ($t_{\textrm{cycl}}\propto \rho^{17/20}T^{-3/2}$) and a decrease in the bremsstrahlung scale ($t_{\textrm{brem}}\propto \rho^{-1}\sqrt{T}$). Thus, even if the cyclotron dominates at the front shock, the bremsstrahlung will become the dominant process lower in the column, inside the high-energy region (see Sect. \ref{cyclotron_simulation}). 
%
%
%

The one-temperature approximation is valid as long as the Coulombian collisions between ions and electrons are sufficient to insure that the two populations have the same temperature despite some energy losses by radiation. When it is not the case, the electrons and ions are thermally decoupled and two-temperature effects appear. The characteristic timescale of electron-ion collisions is given by \citep{Spi65}
\begin{equation}
t_{\textrm{ei}}= 6.22\times 10^{-5} \textrm{ s}\; \left(\frac{\ln \Lambda_{\textrm{C}}}{20}\right)^{-1}\left(\frac{\dot{m}}{1\textrm{ g/cm}^2\textrm{/s}} \right)^{-1}\left(\frac{v_{\textrm{ff}}}{10^8\textrm{ cm/s}} \right)^{4}, \label{tei}
\end{equation}
where $\ln \Lambda_{\textrm{C}}$ is the Coulomb logarithm. These effects are significant when $t_{\textrm{ei}}$ is greater than the timescale of the most efficient cooling process. In the accretion column case, it happens when $t_{\textrm{ei}}\sim t_{\textrm{cycl}}$ (\emph{Regime 3} in Fig.\ref{busschaert_regime}).  
A critical magnetic field can be estimated, as for the previous processes, and is given by
\begin{eqnarray}
B \simeq B_0  \left(\frac{\ln \Lambda_c}{20} \right)^{\frac{20}{57}} \left(\frac{\dot{m}}{1\textrm{ g/cm}^2\textrm{/s}} \right)^{\frac{37}{57}} \left(\frac{v_{\textrm{ff}}}{10^8\textrm{ cm/s}} \right)^{\frac{-157}{57}} 
\left(\frac{S}{10^{15}\textrm{ cm}^2} \right)^{\frac{17}{114}}\label{Bcrit}
\end{eqnarray}
with  $B_0 = 1.9 \times 10^3$~MG.\\

Finally for extremely low accretion rates, as observed at some epochs in most polars \citep{Cro90}, the flow becomes collisionless ($t_{\textrm{ii}}>t_{\textrm{cycl}}$, where $t_{\textrm{ii}}$ is the ion-ion collision timescale) and no shock is formed at the impact (\emph{regime 4} in Fig.\ref{busschaert_regime}). This regime has been identified observationally among different polars referred to as  LARPs (low accretion rate polars, see \cite{Sch02}).  By using the expression of $t_{\textrm{ii}}$ given in \cite{Kyl82}, this specific regime turns out to be crucial when the magnetic field is above a critical value that has the same dependency on the accretion rate, free-fall velocity, and accretion surface as in \eqref{Bcrit} but with a different value of the constant $B_0 = 8.5 \times 10^3$~MG. \\

Considering more typical conditions commonly found in polars (see Paper I), we focus our numerical studies on the first two regimes (\emph{Regimes 1 and 2}), which are the only ones considered in the following sections.  

\subsection{Radiation hydrodynamic numerical code HADES}

 The post-shock region is structured by the cooling processes, thus the position of the reverse shock can be evaluated using Equation \eqref{xs}. For typical values of density and velocity of the incoming flow, $x_{\textrm{s}} \sim  3\times10^7$~cm considering  bremsstrahlung cooling alone\footnote{Adding the cyclotron contribution would induce a shorter cooling timescale, hence a shorter column.}. As a consequence, the curvature of the magnetic field line close to the pole can be neglected, and the accretion column can be assimilated to a cylinder. Moreover, considering the spatial extension of the column, the magnetic field gradient can be neglected along the post-shock region, and the terms linked to the magnetic field in the ideal MHD description vanished so an hydrodynamical description is sufficient to model the base of the column \citep{Cro90, War95, Wu00}.\\

The code \textsc{Hades} is dedicated to the resolution of radiative hydrodynamics equations \citep{Mic11}. Considering our system, we treat radiative losses as a source term in the energy conservation equation 
\begin{eqnarray}
\frac{\partial \rho}{\partial t } + \nabla . (\rho \mathbf{v}) = 0 \; , \label{Euler_masse}\\
\frac{\partial }{\partial t} (\rho \mathbf{v}) + \nabla . \left(\rho \mathbf{v} \otimes \mathbf{v} + P.\bar{\bar{I}}\right) =  \mathbf{0} \;, \label{Euler_impulsion} \\
\frac{\partial}{\partial t} \left( \rho e + \frac{1}{2}\rho v^2 \right) + \nabla .  \left(\rho  e \mathbf{v}  + \frac{1}{2} \rho \mathbf{v} v^2 + P \mathbf{v} \right) =-  \Lambda(\rho,T)\;, \label{Euler_energie}
\end{eqnarray}
where $t$, $\rho$, $P$, $\mathbf{v}$, $e$, $\Lambda(\rho,T), $ and $\bar{\bar{I}}$ are the time, density, pressure, velocity, internal energy
per unit of mass, cooling function of the system, and the identity matrix, respectively. Equations \eqref{Euler_masse}, \eqref{Euler_impulsion}, and \eqref{Euler_energie} come from the conservation of the mass, impulsion, and energy of the system, respectively. \\

The perfect gas approximation is assumed to close the system of equations. The code is designed to solve high-Mach number flows ($\mathcal{M} > 10$) thanks to a MUSCL-Hancock scheme \citep{VLe79} using a HLLE solver to compute the Riemann problem \citep{Ein88}. The cooling function $\Lambda$ is expressed as a sum of the power laws of density and of pressure, thus modelling the various cooling processes at stake in the shocked medium \citep{Bus13_HEDP}. In this numerical model, the initial conditions are the mass density ($\rho_0$), the free-fall velocity ($v_{\textrm{ff}}$), and the temperature of the infalling flow, respectively. In the presented simulations, the infalling flow velocity evolves from $3 \times 10^8$~cm/s ($M_{\textrm{WD}} = 0.4$~$\textrm{M}_{\odot}$) to $5.5 \times 10^8$~cm/s ($M_{\textrm{WD}} = 0.8$~$\textrm{M}_{\odot}$), and the infalling flow temperature has been set at $T = 10^7$~K to ensure a Mach number in the flow higher than ten to obtain a strong shock but also to ensure numerical robustness since the code cannot sustain Mach numbers that are too high. \\

Moreover, a specific boundary condition has been developed to mimic the accretion onto the white dwarf at the bottom of the column. Indeed, since the cooling processes stop when the plasma is sufficiently cooled, a layer of cold and dense material accumulates at the white dwarf surface. On the astrophysical scale, this matter is assimilated by the star, so to avoid this unphysical accumulation, the mass flow is conserved along the column.  Therefore at the bottom of the column, the boundary condition satisfies $\rho_{max} \times v_{min}= \rho_0 \times v_{\textrm{ff}}$. The parameter $v_{min}$ is determined so that the velocity contrast does not exceed three orders of magnitude, insuring the calculation's robustness. We verified that the minimum temperature reached resembles temperature at which the cooling is no longer efficient. More details can be found in \cite{Bus13_HEDP} concerning the implementation and validation of this boundary condition. 

\section{Non-stationary phenomena in the post-shock region}
\label{non-stationnaire}
The structure and dynamics of the PSAC can be strongly modified by unsteady phenomena. In particular the cooling instability \citep{Che82,Mig05} can develop, and the post-shock matter can radiate so strongly that it
again becomes supersonic close to the white dwarf surface, which leads to the formation of a secondary shock \citep{Fal81}.  These unsteady phenomena imply that no stationary regime can be established. These time variabilities are used to explain observations of QPOs in the optical luminosity curves of some polars. The effects of these two mechanisms onto astronomical observations are discussed. Our aim is to quantify the impact of these effects to evaluate the relevancy of the stationary model. These conditions are critical since they are used to determine the white dwarf mass \citep{Wu95}. 

\subsection{The cooling instability}
Thanks to the bremsstrahlung cooling, the post-shock region can developed the \emph{\emph{cooling instability}}  \citep{Che82}. Indeed, according to the Field criterium \citep{Fie65}, a radiating fluid characterized by a cooling function in the form $\Lambda\propto \rho^{2}T^{\beta}$  is instable if $\beta<1$. The linear analysis has been the subject of various studies where the stationary solution is perturbated. The non-linear regime can be explored in details with numerical hydrodynamical simulations \citep{Ima84, Mig05}. \\

\begin{figure}[!ht]
\center
\includegraphics[width=\linewidth]{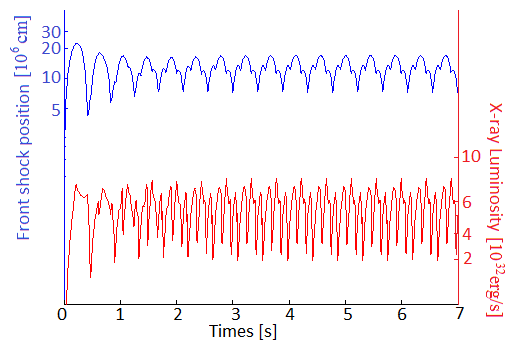}
\caption{Evolution of the front shock position measured from the white dwarf surface (top blue line) and the X-ray ($0.5$ - $10$~keV) luminosity  (bottom red line), for a PSAC with an accretion rate of $\dot{m} =10$~g/cm$^2$/s onto a $0.8\textrm{ M}_{\odot}$ white dwarf and a pure bremsstrahlung cooling.}\label{busschaert_xsLX}
\end{figure}

Figure \ref{busschaert_xsLX} shows the dynamics of the accretion shock simulated using the \textsc{Hades} code with a cooling function $\Lambda\propto \rho^{2}T^{1/2}$, relevant for a pure bremsstrahlung cooling (see Eq. 4). 
The position of the front shock with time is plotted, showing a quasi-periodic oscillation with a period of the order of t$_{\textrm{brem}}$ (see Eq. 5), in accordance with previous works (Mignone 2005). This shock height variation induces a modification of the post-shock volume of the hot and highly radiating material, as well as a modification of the post-shock temperature. As a consequence, the integrated luminosity is also modified. \\
%
%
%

The X-ray luminosity and spectrum are computed using the local bremsstrahlung emissivity combined with the density and temperature profiles extracted from the simulations. To compare with observations, the integrated X-ray luminosity is evaluated by integrating the radiative losses both over the volume of the column and over photon energies between 0.5 and 10 keV that correspond to the XMM-Newton detector range used in Paper I. 
The resulting integrated X-ray light curve is also shown in Fig.\ref{busschaert_xsLX} and clearly displays oscillations at the same main frequency than those observed for the shock height but with a more complex structure. 

\subsection{Formation of secondary shock: Falle criterion}

\begin{figure}[!ht]
\center
\includegraphics[width=\linewidth]{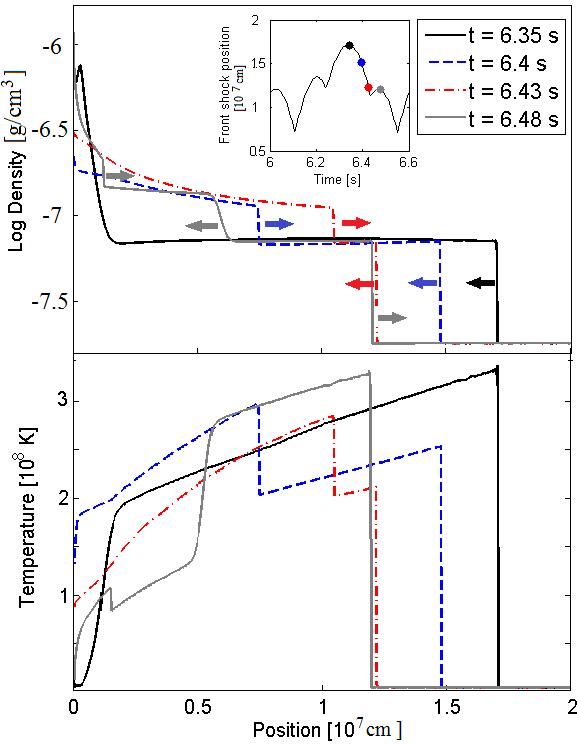}
\caption{Density (top) and temperature (bottom) profiles of the column at successive times across the oscillation (shown in the inset) for an accretion rate of $\dot{m} =10$~g/cm$^2$/s, a white dwarf of $0.8$~M$_{\odot}$, and a pure bremsstrahlung cooling. Starting after t = 6.35 s, a secondary shock clearly forms at the basis of the accretion column and propagates upstream while the main shock falls towards the white dwarf surface.} \label{busschaert_profil_rho}
\end{figure}

The careful examination of the temperature and density profiles of the postshock region and their evolution with time clearly show the formation of a secondary shock (see Fig. \ref{busschaert_profil_rho}). This secondary shock, which has been somewhat overlooked by previous works, is associated with the development of the \emph{c\emph{atastrophic}} instability, or Falle instability, that can take place for a post-shock region cooled mainly by the bremsstrahlung process. This instability was first evoked by \cite{Fal81} in the context of the filamentation of supernovae remnants. Secondary shocks in the post-shock region can arise since right below the main shock, the fluid velocity is subsonic, but owing to the cooling, the flow can again become supersonic close to the white dwarf surface. This kind of phenomenon has already been noted in several studies concerning accretion shocks \citep{Blo89, Str95, Wol89, Mig05}, but its impact onto the main shock dynamics or onto the observables has not been discussed yet. \\

To determine the favourable conditions for developing the catastrophic cooling instability, a local characteristic cooling length, $L_{\textrm{cool}}$, is defined as the product of the sound velocity, $c_{\textrm{s}}$, and the cooling time scale, $t_{\textrm{cool}}$ as 
\begin{equation}
L_{\textrm{cool}}= c_{\textrm{s}} \times t_{\textrm{cool}} \propto T^{3/2-\beta} \rho^{1-\alpha}\;, 
\end{equation} 
where a general cooling function $\Lambda = \Lambda_0 \rho^{\alpha}T^{\beta}$ has been considered. Then, an acoustic wave propagates on a length of the order of $L_{\textrm{cool}}$ before its dynamics is strongly impacted by the cooling processes. It can be shown that to develop the catastrophic cooling instability, the local cooling length has to be smaller than the local distance from the surface of the white dwarf \citep{Fal81}
\begin{equation}
\frac{L_{\textrm{cool}}}{x} < 1\;, \label{Falle1}
\end{equation}
and considering a cooling function characterized by a parameter $\alpha=2$, as for the bremsstrahlung process, then a second criterion is for the second parameter to satisfy \citep{Fal81}
\begin{equation}
\beta < 3/2 \;.\label{Falle2}
\end{equation}

\begin{figure}[!ht]
\center
\includegraphics[width=\linewidth]{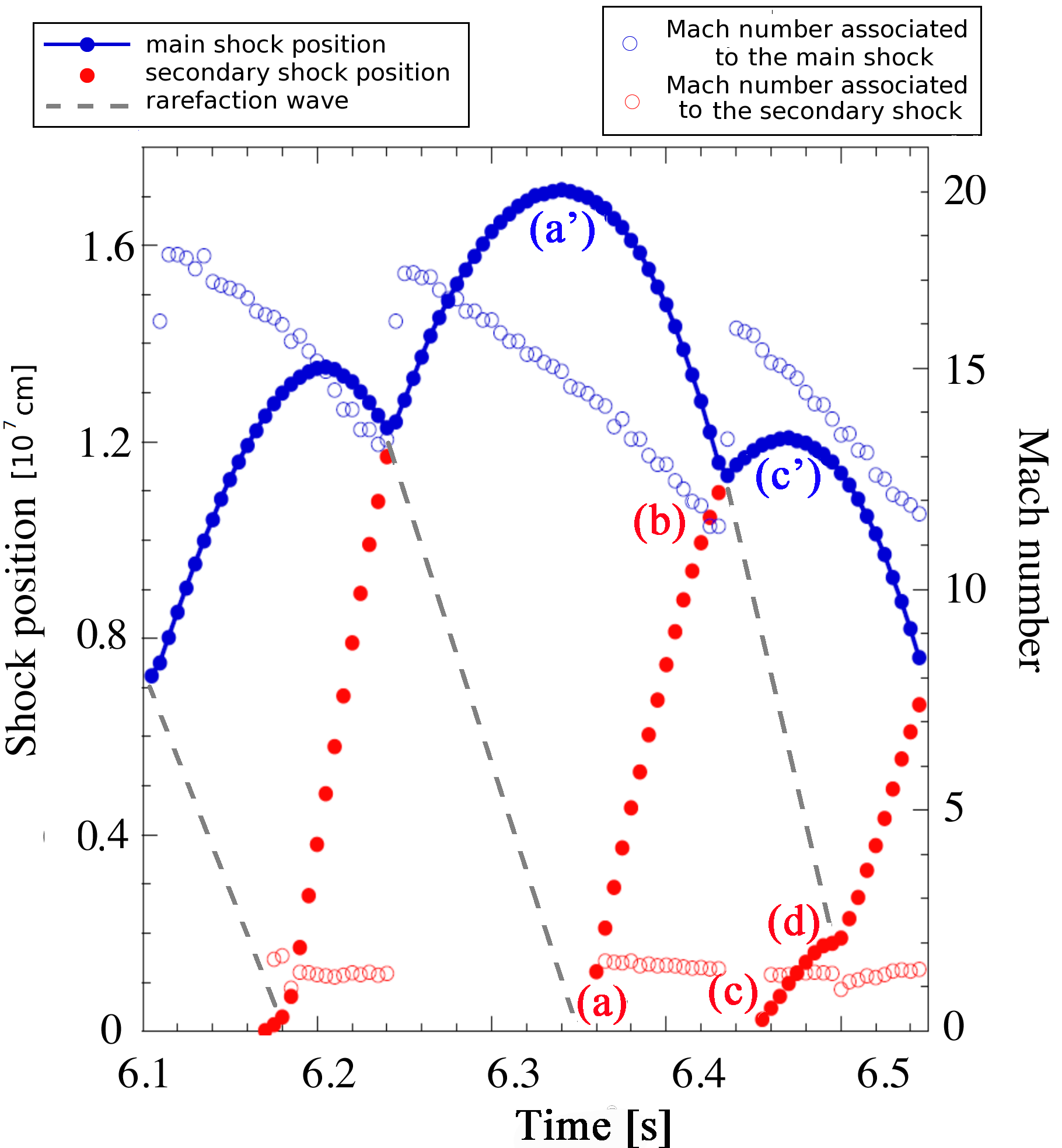}
\caption{Evolution with time of the position of the main shock (blue filled points), the secondary shock (red filled points) and the rarefaction wave (grey dashed line) in the same condition as in Fig.~\ref{busschaert_profil_rho}. The Mach number value for the two shocks are also shown by the open symbols. Different critical times through the oscillation are labelled by letters and discussed in the text.}
\label{busschaert_xs_et_xescondaire}
\end{figure}

The evolutions of density and temperature in the post-shock region are presented at four representative times in Fig.~ \ref{busschaert_profil_rho}. Considering a system cooled by the bremsstrahlung process, the condition \eqref{Falle2} is satisfied since in this case $\beta = 1/2$. The first condition is more complicated to check out. We calculated the ratio  ${L_{\textrm{cool}}}/{x}$ over the column, which decreases from the shock towards the surface. When the secondary shock appears, the ratio tends towards $1$ by superior value close to the surface. To ensure that the structure observed is truly a shock, we computed the Mach number associated with this structure over an oscillation (see Fig.~\ref{busschaert_xs_et_xescondaire}) and checked that the value is around $1.2-2$. Thus, assuming a system cooled only by bremsstrahlung, a secondary shock appears in the post-shock region (see Fig.\ref{busschaert_profil_rho}, $t = 6.4$~s). The secondary shock is formed near the surface of the white dwarf and progresses upstream up to the main reverse shock (see Fig.\ref{busschaert_profil_rho}, $t = 6.43$~s) until the two shocks collide. \\
%
%
%
%
%

The positions of the main and secondary shocks have been tracked and are presented in Fig.~\ref{busschaert_xs_et_xescondaire}, together with the Mach number value for the two shocks. The interaction of the two shocks across the oscillation can be fully described. At maximum extension, the main shock progressively slows down and then falls back towards the white dwarf surface due to the cooling instability (Fig.~\ref{busschaert_xs_et_xescondaire}, a'). Meanwhile, the secondary shock develops from the white dwarf surface (Fig.~\ref{busschaert_xs_et_xescondaire}, a) and propagates upwards until it collides with the main shock (Fig.~\ref{busschaert_xs_et_xescondaire}, b). After both shocks have collided, the main shock propagates upstream, and a rarefaction wave \citep{Zeldo} propagates through the shocked medium towards the white dwarf surface, as expected  (see Fig.\ref{busschaert_profil_rho}, $t = 6.48$~s and Fig.~\ref{busschaert_xs_et_xescondaire}, c'). In the meantime, a new secondary shock has formed close to the white dwarf surface (Fig.~\ref{busschaert_xs_et_xescondaire}, c). On the temperature profile (see Fig.~\ref{busschaert_profil_rho}), both the rarefaction wave and the secondary shock are clearly visible at $t=6.48$~s near $x\sim 5-6 \times 10^6$~cm and $x\sim 2 \times 10^6 $~cm, respectively. As the rarefaction wave encounters the newly formed secondary shock (see Fig.~\ref{busschaert_xs_et_xescondaire}, d), the shock accelerates further and then collides again with the main shock. Then this cyclic behaviour starts again. The secondary shock always appears close to the white dwarf surface when the extension of the PSAC is close to its maximum. After each collision between the two shocks, only the main shock remains and is pushed away from the white dwarf surface. \\

Secondary shocks are also present in previous studies, which considered at least four different boundary conditions (see Fig. 5 in \cite{Mig05}), though such shocks were not fully discussed in this work. To ensure that this physical phenomena truly exists, we verified that its presence and this cyclic behaviour are independent of boundary conditions at the basis of the accretion column and initial conditions.   \\

To study the development of the catastrophic cooling instabilities, several cooling functions have been tested with the parameters $\alpha =2$ and various $\beta$ values. For values of $\beta$ higher than $0.7$, this instability does not develop. Indeed, even if the second criterion (see Eq.\eqref{Falle2}) is fulfilled, the cooling process acts too quickly after the initial collision to allow the main shock going far enough from the white dwarf surface to fulfil the first criterion (see Eq.\eqref{Falle1}).  Also, it should be noted that when both cyclotron and bremsstrahlung are effective, considering the stabilizing effect of cyclotron losses, the secondary shock does not appear for a high magnetic field for the same reason. \\ 

\begin{figure}[!ht]
\center
\includegraphics[width=\linewidth]{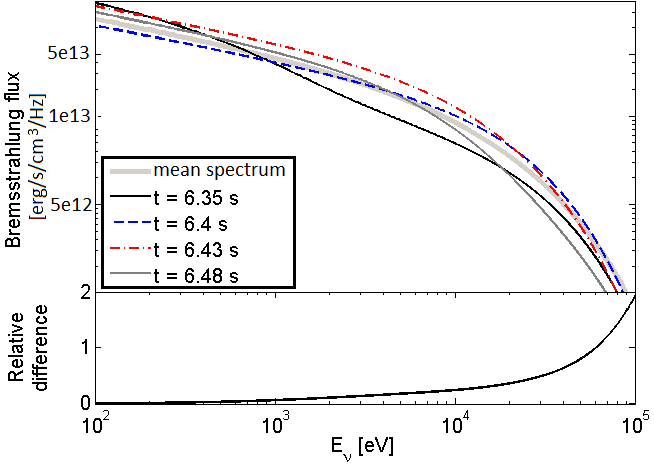}
\caption{Top: Evolution of the emitted bremsstrahlung spectrum through the oscillation for the same parameters and the times indicated in the inset of Fig.~\ref{busschaert_profil_rho}. The mean spectrum calculated over a couple of oscillations is represented by the thick grey line. Bottom: Relative difference between the mean spectrum and the spectrum emitted by the theoretical stationary column.}
\label{evolution_spectre}
\end{figure}

From the simulations, the detailed X-ray spectrum emitted by the PSAC over an oscillation can be computed and is shown in Fig.~\ref{evolution_spectre}. The spectrum evolves strongly through the different phases of the oscillation. When only the primary shock is present ($t=6.35$~s), the emitted spectrum is softer than when the secondary shock has appeared ($t = 6.4-6.43$~s). The relative difference between the mean spectrum calculated over a couple of oscillations and the spectrum emitted by the theoretical stationary column is presented at the bottom of Fig.~\ref{evolution_spectre}. The difference between the two spectra is quite negligible at low energy, but at high energy, the mean spectrum is significantly lower than the stationary one. Thus, if we attempt to evaluate the white dwarf mass through the mean spectrum, we expect to find a lower mass. \\

%
%
%
Considering both cooling instabilities, the column structure, when cooled mainly by the bremsstrahlung process, is quite different from the steady state solution. This strongly affects both light curves and emitted spectra.

\section{Influence of system parameters on luminosity variations}
\label{parametre}

In Paper I, observational X-ray light curves of a set of polars were analysed to search for oscillations as predicted by the cooling instability model. To directly compare these observational constraints with the theoretical expectations from the shock model, we used the \textsc{Hades} simulations to compute the amplitudes and frequencies of the expected oscillations more precisely for different sets of parameters that are relevant to the polars, considering both bremsstrahlung and cyclotron cooling.

\subsection{Bremsstrahlung dominated columns}

\begin{figure}[!ht]
\center
\includegraphics[width=\linewidth]{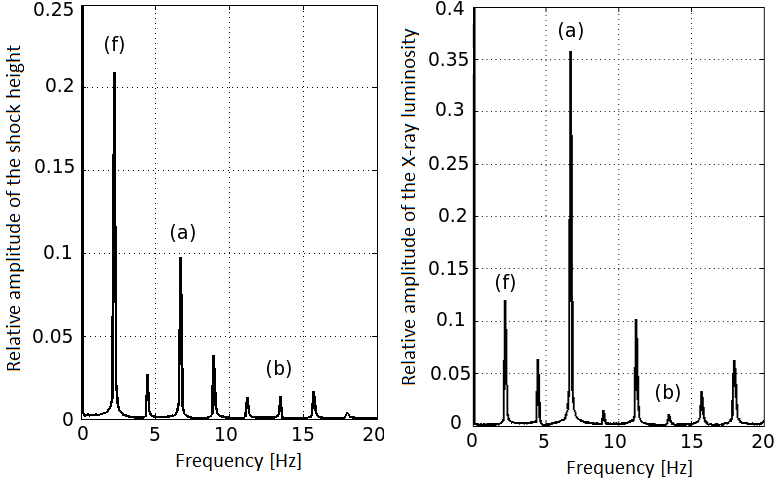}
\caption{Fast Fourier transform of the front shock position (left) and the X-ray luminosity (right) respectively normalized by their mean values for the same parameters as in Fig.~\ref{busschaert_profil_rho}. The frequencies are identical for both signals but with different amplitudes. The main modes discussed in the text are referred as (f), (a), and (b). }
\label{busschaert_fftxsLX}
\end{figure}

The temporal characteristics of the front shock position and the computed PSAC X-ray flux emerging from the simulations were analysed using fast Fourier transforms (FFT) to better understand the properties of the quasi-periodic phenomena. Representative power spectra of the shock front position and X-ray luminosity are presented in Fig.~\ref{busschaert_fftxsLX} where we consider a white dwarf of $0.8$~M$_{\odot}$ with an accretion rate of  $\dot{m} = 10$~g/cm$^2$/s and only taking the bremsstrahlung losses into account. The FFT covers simulation data obtained at 0.01 s resolution over 2048 time steps at long times between $t = 19.76$~s and $t = 30$~s.  \\

Several discrete frequencies emerge from theses analyses, and the same frequencies are present in both signals but with different relative amplitudes. The different amplitudes are explained by considering that not only does the variation in the emission volume modify the luminosity but also the physical conditions of temperature and density throughout the post-shock region. For instance, the post-shock temperature depends on the pre-shock velocity in the shock frame. Thus, when the shock is going upstream (resp. downstream), the infalling flow goes faster (resp. slower) than the free-fall velocity, and the post-shock temperature is higher (resp. lower). This phenomenon is clearly visible in  the temperature profiles presented in Fig.~\ref{busschaert_profil_rho}. Amongst the most significant frequencies, the lowest is found  at 2.24~Hz (period of 0.45~s) (mode \emph{f}). This frequency is very visible in the first instants (see Fig.~\ref{busschaert_xsLX}) and remains present later. It is correlated with the setting up of the PSAC structure: when the flow impacts the surface, the shock goes away from the surface and progressively slows down. At $t=0.24$~s, the shock falls back and reaches the minimum position at $t=0.42$~s before bouncing against the secondary shock. The second strongest mode is present at a frequency of 6.7~Hz (period of 0.15~s) (mode \emph{a}). This typical timescale corresponds to the cooling timescale that is calculated using Eq. \eqref{t_brem}. Modes of higher frequencies are also present and can make  a significant contribution to the oscillations. A mode (b) has been emphasized in Fig.~\ref{busschaert_fftxsLX}, since  it has a strong contribution for higher magnetic field strength. For the regime with pure bremsstrahlung cooling, the FFT exhibits the same overall behaviour for the different white dwarf masses tested ($M_{\textrm{WD}}=0.4, 0.6, 0.8$~M$_{\odot}$). However, the relative contribution of each mode can be slightly different, and it can be demonstrated how the frequency depends on the accretion rate and the white dwarf mass. \\

%
%
\begin{figure}[!ht]
\center
\includegraphics[width=\linewidth]{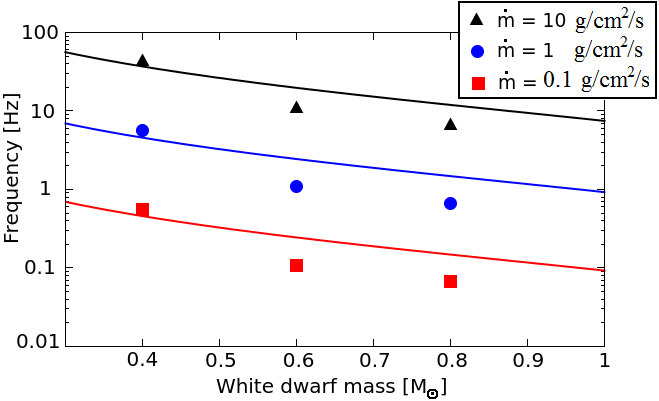}
\caption{Variation in the main oscillation frequency (mode a) of the X-ray light curve as a function of the white dwarf mass, for different accretion rates in the case of pure bremsstrahlung cooling. The curves represent the best fits following the theoretical tendency given by the relation \eqref{tendance}.} 
\label{busschaert_frequenceMmdot}
\end{figure}

%
Considering the bremsstrahlung cooling as dominant (see Regime 1, Fig.~\ref{busschaert_regime}), the mechanism that induces the oscillations is the cooling instability with the oscillation frequency expected to evolve as the inverse of the cooling time (see Eq.~\eqref{t_brem})
\begin{equation}
\nu  \propto t_{\textrm{brem}}^{-1} \propto \dot{m} / v_{\textrm{ff}}^2 \;.
\end{equation}
Introducing the expression of the free-fall velocity coupled to the white dwarf mass-radius relation \citep{Nau72}, the frequency depends on the specific accretion rate and the white dwarf mass as  
\begin{equation}
\nu = A_0 \dot{m} \left[ \left( \frac{1.44 \textrm{M}_{\odot}}{M_{\textrm{WD}}} \right)^{8/3} - \left( \frac{1.44 \textrm{M}_{\odot}}{M_{\textrm{WD}}} \right)^{4/3} \right]^{1/2} \;,\label{tendance}
\end{equation}
where $A_0$ is a constant. In Fig.~\ref{busschaert_frequenceMmdot}, we represent the dominant oscillation frequency of the X-ray luminosity (mode a), extracted from simulations for three different white dwarf masses and accretion rates. The dependence from Eq. \eqref{tendance} is plotted as solid lines and gives a good estimate of the oscillation frequency. The constant $A_0$, which provides the best fit, is $A_0 = 0.9, 0.9,$ and $0.7$~cm$^2$/g for $\dot{m}=0.1, 1,$ and $10$~g/cm$^2$/s,
respectively, while the expected theoretical value is $A_{0,th} = 0.5 $~cm$^2$/g. These values are slightly different since the theoretical value is evaluated just behind the shock and does not take the strong variations along the column into account. \\

These results imply that, when QPOs are observed, their frequencies should allow constraining the value of the accretion rate and white dwarf mass (see Paper I). \\

%
%
%
\subsection{PSAC efficiently cooled by the cyclotron process}
\label{cyclotron_simulation}

\begin{figure}[!ht]
\center
\includegraphics[width=\linewidth]{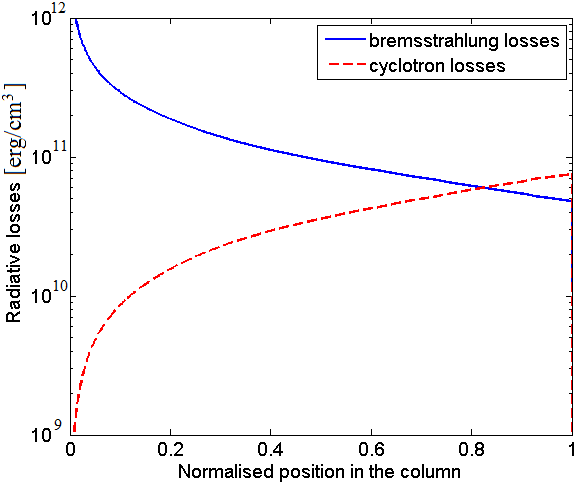}
\caption{Variation in the bremsstrahlung (solid blue line) and cyclotron (dotted red line) losses along the column assuming a white dwarf of $0.8$~M$_{\odot}$, an accretion rate of $\dot{m} =10$~g/cm$^2$/s and $\epsilon_{\textrm{s}}=1.2$ (corresponding to a magnetic field of $30.3$~MG) at $t=100$~s.  The abscisses have been normalized to the front shock position ($x_{\textrm{s}} \sim 7.7 \times 10^6$~cm).}\label{busschaert_perte}
\end{figure}

When the magnetic field is strong enough, the cyclotron cooling becomes efficient  and a two-process cooling has to be considered. The losses for the bremsstrahlung and cyclotron processes are presented in Fig.~\ref{busschaert_perte} as a function of the position in the column for a typical configuration. The cyclotron losses are higher than the bremsstrahlung ones at the shock front in this case since $\epsilon_{\textrm{s}} = 1.2,$ which corresponds to a magnetic field of $30.3$~MG, assuming a column section of $10^{15}$~cm$^2$. Because the bremsstrahlung mechanism is more sensitive to variation in density than temperature (see Eq.\, \eqref{brem}), the losses progressively increase from the shock to the surface of the white dwarf along with the density. In contrast, since cyclotron mechanism is more sensitive to variations in temperature than in density (see Eq.\, \eqref{cyc}), it is more efficient at the shock where matter is hotter than closer to the white dwarf surface. Thus, $\epsilon_{\textrm{s}}$, defined as the ratio of cooling times (see Eq.\, \eqref{epsilon_s}), is always maximum at the shock and then decreases along the column towards the white dwarf surface. \\

%
%
%
The analysis in Paper I gives upper limits for the amplitude of quasi-periodic oscillations in X-rays. Up to now, oscillations in light curves have only been observed in the optical range for a few selected sources. 
In the context of the shock instability model for polars, it should be emphasized that optical variations associated to cyclotron are expected together with the X-ray ones linked to bremsstrahlung. 
From the simulations, we extracted both components separately and compare them here in terms of amplitude and frequency. \\

\begin{figure}[!ht]
\center
\includegraphics[width=\linewidth]{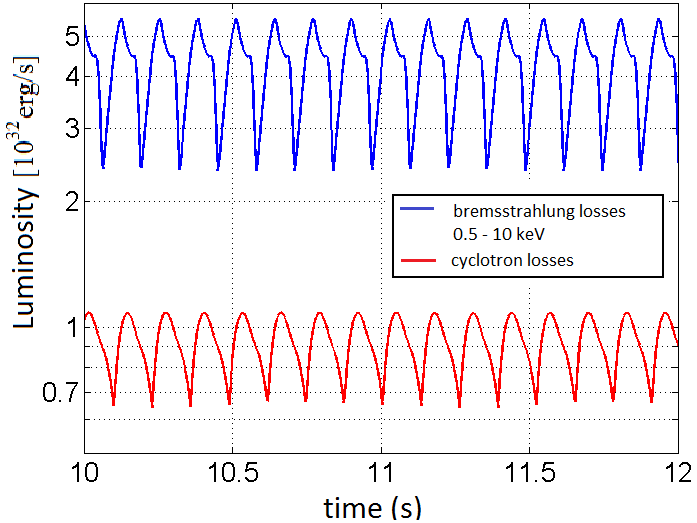}
\caption{Variation in the (0.5-10 keV) bremsstrahlung (upper blue line) and cyclotron (lower red line) luminosity considering a white dwarf of $0.8$~M$_{\odot}$, an accretion rate of $\dot{m} =10$~g/cm$^2$/s, and a magnetic field of $18.7$~MG ($\epsilon_{\textrm{s}}=0.3$).}
\label{busschaert_LXLcyc}
\end{figure}

\begin{figure}[!ht]
\center
\includegraphics[width=0.8\linewidth]{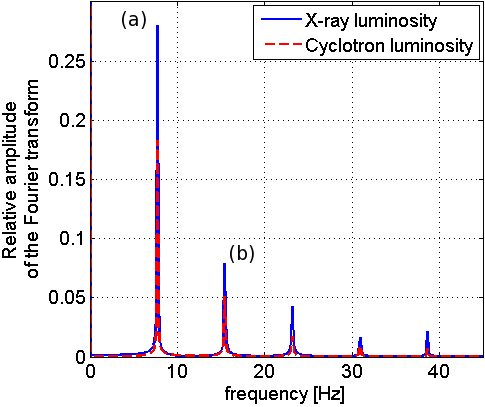}
\caption{Fourier transform of the (0.5-10 keV) bremsstrahlung luminosity and cyclotron total luminosity with the same parameters as in Fig.~\ref{busschaert_LXLcyc}.}
\label{busschaert_fftLXLcyc}
\end{figure}

In Fig.~\ref{busschaert_LXLcyc}, the luminosity in the optical range due to the cyclotron losses is compared to the X-ray emission due to the bremsstrahlung process, considering a white dwarf with $M_{\textrm{WD}} = 0.8 $~M$_{\odot}$ and an accretion rate of $10$~g/cm$^2$/s, for a typical magnetic field of $18.7$~MG ($\epsilon_{\textrm{s}} = 0.3$). The behaviour in the two energy ranges is very similar. We present in Fig.~\ref{busschaert_fftLXLcyc} the corresponding Fourier transforms of the cyclotron and bremsstrahlung emissions, where the power shown is the oscillation amplitude normalized by the mean luminosity. They present the exact same frequencies of oscillations but with an oscillation amplitude in the dominant mode (a), which is lower for the cyclotron signal ($\sim 18\%$) than the bremsstrahlung one ($\sim 28\%$). This behaviour has been systematically observed in all simulations. Thus, we expect to observe stronger oscillations in the X-ray light curve than in the optical domain. As already noted by \citet{Wu92}, we confirm that the two signals are slightly out of synchronization.  \\

%
%
%
%
%
%
Considering the perturbative analysis, the system is expected to stabilize as cyclotron becomes a strong contribution in the radiative losses \citep{Sax98, Sax, Sax99}. In the results presented in this work, for a white dwarf of $0.8$~M$_{\odot}$, the oscillations have been considered as completely damped when the front shock oscillation is only a few meshes in length. It corresponds to a luminosity variation of less than $\sim 0.1 \%$. We observed that for the accretion rate of $\dot{m} \sim 1$ ~g/cm$^2$/s, the oscillations are effectively damped for $\epsilon_{\textrm{s}} \sim 1.5$; and for  $\dot{m} \sim 10$ ~g/cm$^2$/s, the oscillations are almost damped for the same value of $\epsilon_{\textrm{s}}$. \\

\begin{figure}[!ht]
\center
\includegraphics[width=\linewidth]{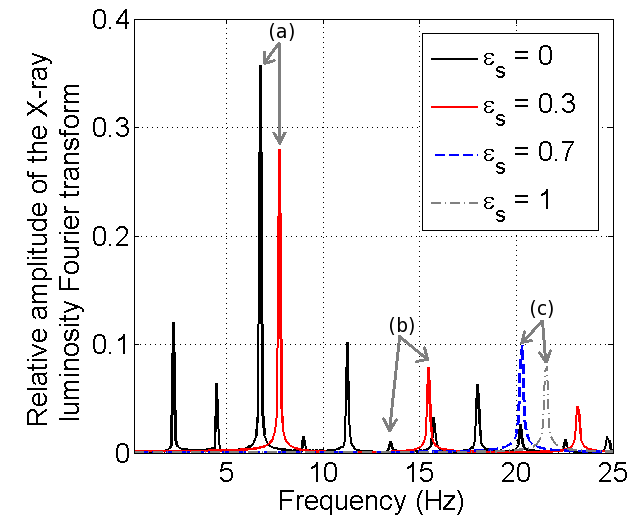}
\caption{Changes in the FFT of the X-ray light curve for \textbf{a growing} magnetic field characterized by $\epsilon_{\textrm{s}}$, for a white dwarf of $0.8~\textrm{M}_{\odot}$ and an accretion rate of $\dot{m} =10$~g/cm$^2$/s. The range in $\epsilon_{\textrm{s}}$ from 0 to 1 corresponds to a magnetic field from 0 to $28$~MG. The different frequency modes  are identified as (a), (b), and (c) and listed in Table~\ref{table_mdot10}.}
\label{busschaert_fftB}
\end{figure}

As the magnetic field strength increases, the low-order modes are expected to be rapidly damped. This type of behaviour is found in our numerical simulations. The typical evolution of the X-ray light curve FFT is presented in Fig.~\ref{busschaert_fftB} for various values of $\epsilon_{\textrm{s}}$ that correspond to a magnetic field strength ranging  from $0$ to $28$~MG, considering an $M_{\textrm{WD}} = 0.8$~M$_{\odot}$ white dwarf and an accretion rate of $\dot{m} =10$~g/cm$^2$/s. For weak magnetic fields, the dominant mode is around ($\nu \simeq 6-7$~Hz), but for higher fields, this mode decreases strongly and disappears, while higher modes ($\nu \simeq 20-25$~Hz) appear. \\

%
%
\begin{figure}[!ht]
\center
\includegraphics[width=\linewidth]{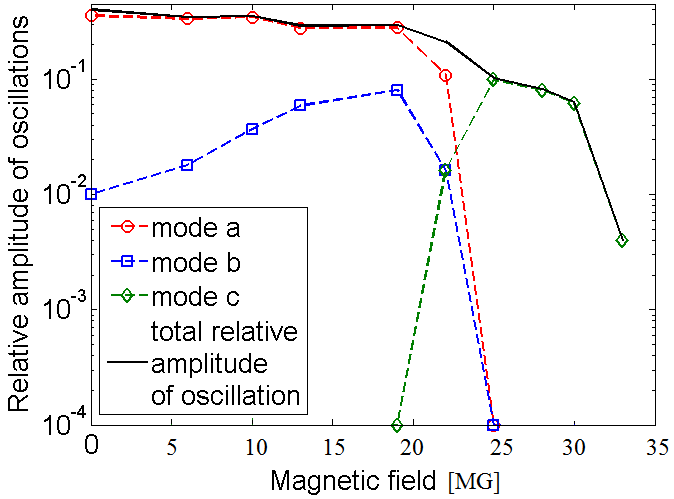}
\caption{Variation in the relative amplitude of the X-ray oscillations for the three main modes observed in the FFT for a white dwarf of $0.8$~M$_{\odot}$ and an accretion rate of $\dot{m} =10$~g/cm$^2$/s. The total relative amplitude is given by the full black line. }\label{busschaert_fig9}
\end{figure}

%
In Fig.~\ref{busschaert_fftB}, it is clear that the distribution of the various modes and their identification are not obvious. However, a few main modes can be tracked when the magnetic field grows. The frequencies and relative amplitudes of the three main modes are reported in Tables~\ref{table_mdot1} and \ref{table_mdot10} and plotted in Fig.~\ref{busschaert_fig9}. We see that Mode (a) decreases monotonically as the magnetic field increases. Its amplitude becomes negligible for a magnetic field around $23$~MG. Mode (b) increases initially and then drops suddenly around the same critical magnetic field as Mode (a). Mode (c) has the same behaviour as Mode (b), but its amplitude keeps increasing when the first two modes decrease strongly. Its amplitude ultimately decreases for higher magnetic field amplitude. The different behaviours exhibited by these first three modes is interesting since it reveals the great richness of the non-linear dynamics of the PSAC. Moreover, it should be emphasized that progressively, as the magnetic field grows, the power is distributed in higher order modes. It should be noted that this non-linear behaviour is coherent with the linear stability studies published by \cite{Sax98}. In Tables~\ref{table_mdot1} and \ref{table_mdot10}, the total relative amplitude is also given as a quadratic sum of the amplitudes in the different modes when the power is split into different frequencies. \\

%
%
%
For a given mode, a frequency variation with the magnetic field strength is observed in the simulations that can be explained analytically. Under the assumption that oscillations are induced by the cooling instability, the expected frequency is directly connected to the cooling time by the relation  
\begin{equation}
\nu \propto 1/t_{\textrm{cool}} = \frac{1}{t_{\textrm{brem}}}+\frac{1}{t_{cyc}}=\frac{1}{t_{\textrm{brem}}}\left[1+\epsilon(B)\right]\;,
\end{equation}
where $\epsilon(B)$ is the ratio between the bremsstrahlung and cyclotron cooling time in each spatial position. If we suppose that the typical timescales calculated behind the shock are representative of the value along the accretion column, then $\epsilon(B) \simeq \epsilon_{\textrm{s}}$, and the relation can be written as 
\begin{equation}
\nu = \nu_{brem,s} \left[1+ C_0 \left(\frac{B}{10~\textrm{MG}}\right)^{57/20}\right]\;,\label{nu}
\end{equation}
where $\nu_{brem,s}$ is the oscillation frequency at the front shock in the asymptotic bremsstrahlung dominated regime, and $C_0$ is a constant that depends on the accretion rate and the free fall velocity as
\begin{eqnarray}
C_0 \simeq 2\times 10^{-4} \left[ \frac{S}{10^{15}~\textrm{cm}^2}\right]^{-17/40} \left[ \frac{\dot{m}}{1~\textrm{g/cm}^2\textrm{/s}} \right]^{-37/20} \nonumber \\
\left[ \frac{v_{\textrm{ff}}}{10^8 \textrm{cm/s}}\right]^{117/20} \,. \label{c0}
\end{eqnarray}

The specific behaviour of the oscillation mode frequencies predicted here is different to the results presented by \cite{Wu92} in the context of a stochastic accretion.  In \cite{Wu92}, the suggested simplified evolution law $\nu\propto (1+C_0 B^2)$ is only based on phenomenological assumptions and not justified by physical considerations. \\

\begin{figure}
\center
\includegraphics[width=\linewidth]{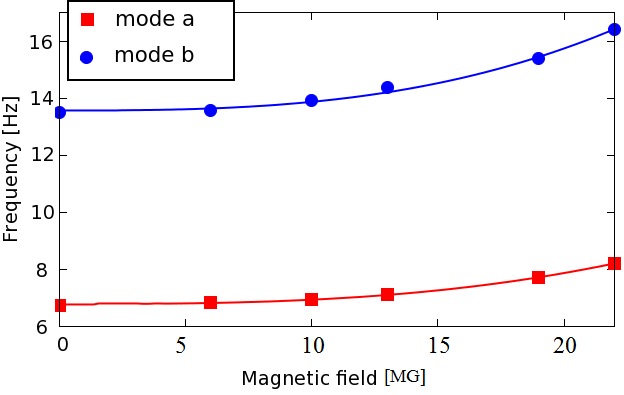}
\caption{Evolution of oscillation frequency with magnetic field for the first two modes (a and b) for a white dwarf of $0.8$~M$_{\odot}$ and an accretion rate of $\dot{m} =10$~g/cm$^2$/s.  The lines are the best fits according to the predicted dependency of Eq. \eqref{nu}.}
\label{evol_frequence_B}
\end{figure}

 The evolution of the frequency for a range of magnetic fields from 0 to $22$~MG is presented in Fig.~\ref{evol_frequence_B} for the two main modes (a) and (b), assuming a 0.8~M$_{\odot}$ white dwarf and an accretion rate of $10$~g/cm$^2$/s. It should be noted that for strongest magnetic fields, the amplitude of those modes decreases drastically so no oscillation frequency can be found. The curves in Fig. \ref{evol_frequence_B} represent the best fits using the analytical trend presented in Eq. \eqref{nu}. The constant determined from these fits is $C_0 = 0.022$~ and is the same for the two modes. In this specific configuration, the expected analytical value of the constant as derived from Eq. \eqref{c0} is $C_{0, \textrm{th}} = 0.061$. The difference with the fitted value from the numerical simulations comes from the structure of the post-shock region. Indeed, to theoretically infer the characteristic timescale, only the values just behind the shock have been taken into account (see also 4.1).

%
%
%

\section{Conclusions}

In this work, a numerical study of the dynamics of PSAC has been presented. Radiation hydrodynamic simulations have shown that the coupling between magnetic fields, radiation processes, and extreme hydrodynamics leads to a complex dynamics and a wide variety of dynamical behaviours. To validate the numerical simulations, they have been compared to theoretical stability criteria, analytical results, and dimensional tendencies. Good agreement has been obtained, in general. To link simulation results to observational data presented in Paper I, a post-treatment of numerical results were performed to extract and study the X-ray and optical luminosities and their dynamics, the X-ray spectral luminosity, and the frequency spectrum. Given the observational data (see Paper I), we have focused our analysis on the two most important regimes of PSAC: the bremsstrahlung-dominated regime (weak magnetic field and/or high accretion rate) and the coupled bremsstrahlung-cyclotron regime (strong magnetic fields and/or lower accretion rate).\\

In the bremsstrahlung-dominated regime, one of the most important results is the development of the catastrophic instability, which leads to the formation of secondary shocks. According to radiation hydrodynamic simulations, those additional shocks modify the dynamics of the primary shock, as well as density and temperature profiles. It induces modifications on the total and spectral emissivities and plays an important role in the oscillations of the shock height. The complex coupling between the secondary shock dynamics and the cooling instability development leads to quasi-periodic oscillations in the light curves. Specifically, the secondary shock implies a harder X-ray spectrum than in the stationary regime, which can affect the mass determination from the classical X-ray approach. The presence and role of the secondary shocks in PSAC have not been studied much as yet. In principle, the presence of the secondary shock can also be revealed through variations induced in the X-ray spectrum across the oscillations. However, the sensitivity of the present X-ray telescopes is not yet high enough to detect such variations. Laboratory experiments, using powerful lasers already in use or near completion such as NIF or LMJ, are a promising way to directly investigate these shocks. \\

In the coupled bremsstrahlung-cyclotron regime, numerical simulations show the stabilizing effect of the cyclotron process. The predicted amplitude of oscillations in the X-ray light curve considering a pure bremsstrahlung regime is $\sim 30\%$. As the magnetic field grows, the amplitude of oscillation decreases and a critical magnetic field depending on the accretion parameters can be found where no oscillation persists. Thanks to numerical simulations, a more detailed study of the distribution of power in the different modes of oscillations has been explored. In particular, the low-order modes are the first to be damped as the magnetic field grows, and the power is then distributed in higher order modes. These two effects have been theoretically expected in the linear regime and persist with non-linear effects. An important correlation between the magnetic field and the oscillation frequency has been established. A comparison between the numerical results and the theoretical relations based on dimensional arguments leads to very good agreement.\\

Numerical simulations exhibit strong oscillations in the X-ray luminosity with the same frequencies as in the optical luminosity curve. They show the presence of QPOs with a wider range of frequencies in the X-ray domain (0.6-50 Hz) than commonly assumed. This wider range, which has never been fully explored observationally until now, has been investigated in Paper 1 with negative results. There is therefore no compelling evidence yet of significant X-ray oscillations, in contrast to what is expected from shock instabilities. These numerical results raise important questions on the origin of physical processes that lead to QPO. \\

The inconsistency between X-ray and optical observational data, on the one hand, and between numerical expectations and observations, on the other, highlights that the physics of QPO is still not completely understood. 
The situation is somewhat similar among classical  T Tauri stars for which shock-heated accreting material is expected to give rise to QPOs  \citep{Kol08, Sac08} that are not yet observed \citep{Dra09, Gun10}. In this context, 2D axisymmetric MHD simulations have shown that significant distortions of the accretion structure can occur as a result of the gas pressure. At the bottom of the column, accumulation of matter may lead to outflows parallel to the star surface that can increase the cooling, and the oscillations are expected to be reduced or suppressed (Orlando 2010, 2013). However, the analogy between polars and T Tauri stars is somewhat limited. Indeed, for the typical magnetic field and velocity and density regimes, the plasma parameter (noted $\beta=P_{th}/P_{mag}$, where $P_{th}$ and $P_{mag}$ are respectively the thermal and magnetic pressure) is significantly different. It is in the range $\beta_{\textrm{TTauri}} \sim 10^{-2}-10^2$ for T Tauri \citep{Matsakos}, but only $\beta_{\textrm{polar}} \sim 10^{-4}-10^{-3}$ for polars. Therefore the magnetic collimation is much more efficient for polars, and 2D effects are expected to be less important. In T Tauri, low $\beta$ values have been shown to lead to a fragmentation of the column into independent oscillating tubes or fibrils \citep{Orl10, Matsakos}. If out of phase, the oscillations of the different fibrils can cancel each other out. Such an effect can also occur in polar PSAC and would also induce a frequency widening, as well as a significant decrease in the overall QPO amplitudes. These effects will be the subject of an upcoming study where numerical simulations will take multi-dimensional effects into account, as well as non-homogeneous accretion and a more detailed treatment of the boundary conditions at the white dwarf photosphere.\\

Additionally, it should be emphasized that the PSAC presents interesting similarity properties that allow us to reproduce a scaled model in laboratory on measured scales \citep{Fal11b, Bus13_NJP}. This is the main objective of a laboratory astrophysics project: the POLAR project \citep{Fal11a}. The possibility of producing relevant experiments has been demonstrated on the LULI2000 facility \citep{Fal12}. The experimental way is a complementary approach to increase our understanding of the radiation hydrodynamics of accretion shock. Thus the QPO physics will be explored using dedicated designed experiments with powerful lasers.

\section*{Acknowledgments}
We wish to thank an anonymous referee for helpful comments.
C. B. and C. M. acknowledge financial support from the PNPS of CNRS/INSU, France, as well as from the Scientific Council of the Observatoire de Paris (France). C. B., E.F., M. M., and C. M. acknowledge financial support from the PNHE, France. The authors also thank H. C. Nguyen and M. Mancini for the development of HADES. 


\begin{onecolumn}
\begin{table}
\caption{Characteristics of the main oscillation modes extracted from the simulations of an accretion column onto a 0.8~M$_{\odot}$ white dwarf with an accretion rate $\dot{m} = 1$~g/cm$^2$/s and for different magnetic field strengths. The first column gives the value of the ratio $\epsilon_{\textrm{s}}$ defined in \eqref{epsilon_s}, the second column gives the corresponding magnetic field strength considering an accretion surface of $S=10^{15}$~cm$^2$, and the third column gives the value of the theoretical stationary shock height. The subsequent columns give for the first three main modes (a, b, and c), the frequency and the relative amplitude of the oscillations in the (0.5-10 keV) X-ray range ($L_{\textrm{X}}$)  and the optical cyclotron flux ($L_{\textrm{c}}$). The last two columns give the quadratic mean value of the relative amplitude integrated over all the oscillation modes, including those not reported in the table.} 
\begin{tabular}{|c|c|c||c|c|c||c|c|c||c |c|c||c|c|}
\cline{4-12} \multicolumn{3}{c|}{} & \multicolumn{3}{|c||}{Mode a}& \multicolumn{3}{|c||}{Mode b}&  \multicolumn{3}{|c|}{Mode c} & \multicolumn{2}{|c}{} \\
\hline
$\epsilon_{\textrm{s}}$&$B$ &$x_{\textrm{s, m}}$   & $\nu$  &  $L_{\textrm{X}}$ & $L_{\textrm{c}}$ & $\nu$    & $L_{\textrm{X}}$ &  $L_{\textrm{c}}$ & $\nu$   &  $L_{\textrm{X}}$ &  $L_{\textrm{c}}$ & $L_{\textrm{X}}$& $L_{c}$\\
&[MG]& [$10^7$ cm]  & [Hz]  &   &  & [Hz]   & &  &  [Hz]  & &  & tot & tot \\
\hline

\hline
 0 & 0 &  10.95 & 0.664 & 0.337 & ...& 1.11 & 0.026 & ... &  &  & ... & 0.362 & ...\\
\hline
0.01 & 1.3  & 10.91 & 0.674 & 0.367 & 0.126 & 1.11 & 0.024 &  0.013 &  & & & 0.376 & 0.168  \\
\hline
0.05 & 2.2  & 10.72 & 0.684 &  0.300 & 0.114 &  1.38 & 0.025 & 0.071 &  & &  &  0.391 & 0.168\\
\hline
0.1 & 2.8 &  10.49  & 0.703 & 0.331 & 0.143 & 1.41 &  0.036 & 0.050 & ... & ... &...  & 0.391 & 0.145\\
\hline
0.3 & 4.2 &  9.72  & 0.781 & 0.126 & 0.058 & 1.55 &  0.040 & 0.012 & 1.85 & 0.059 & 0.008  & 0.185 & 0.110\\
\hline
0.5 & 5.0 &  9.10 &  0.830 &  0.095 & 0.079 & 1.65 & 0.013 & 0.007 & 1.96 & 0.054 & 0.016 & 0.199 & 0.100\\
\hline
0.7 & 5.6  & 8.59 &  0.869 & 0.009 & 0.008 & ... & ...  & ... & 2.041 & 0.141 & 0.043 & 0.143 & 0.044\\
\hline
1 & 6.4  & 7.95 & ... & ... & ... & ... & ... &  ... & 2.18 & 0.122 & 0.044 & 0.123 & 0.045\\
\hline
1.2 & 6.8 & 7.60 &  ... & ... & ... & ... & ... & ... & 2.26 & 0.068 & 0.027 & 0.069 & 0.028 \\
\hline
1.5 & 7.4&  7.15 & ... & ... & ... & ... & ... & ... & ... & ... & ... & ... &  ... \\
\hline
\end{tabular}
\label{table_mdot1}
\end{table}
%
%
\begin{table}
\caption{
Same as Table~ \ref{table_mdot1} except for $\dot{m}=10$~g/cm$^2$/s
} 
\begin{tabular}{|c|c|c||c|c|c||c|c|c||c|c|c||c|c|}
\cline{4-12} \multicolumn{3}{c|}{} & \multicolumn{3}{|c||}{Mode a}& \multicolumn{3}{|c||}{Mode b}&  \multicolumn{3}{|c|}{Mode c} & \multicolumn{2}{|c}{} \\

\hline
$\epsilon_{\textrm{s}}$&$B$ &$x_{\textrm{s, m}}$   & $\nu$  &  $L_{\textrm{X}}$ & $L_{\textrm{c}}$ & $\nu$    & $L_{\textrm{X}}$ &  $L_{\textrm{c}}$ & $\nu$   &  $L_{\textrm{X}}$ &  $L_{\textrm{c}}$ & $L_{\textrm{X}}$& $L_{c}$\\
&[MG]& [$10^7$ cm]  & [Hz]  &   &  & [Hz]   & &  &  [Hz]  & &  & tot & tot \\

\hline
\hline
 0 & 0 & 1.095 & 6.74 & 0.357 & ... & 13.5 & 0.01 & ... & ... & ...  & ... & 0.401 & ... \\
\hline
0.01 & 6 & 1.09   & 6.84 & 0.334 & 0.134 & 13.57 & 0.018 & 0.05 & ... & ...  & ...  & 0.347 & 0.145\\
\hline
0.05 & 10 & 1.07  & 6.93 &  0.347 & 0.15 & 13.9 &  0.037 & 0.062 & ... & ...  & ...  & 0.353 & 0.164\\
\hline
0.1 & 13 & 1.05   & 7.13 &  0.278 & 0.148 & 14.36 &  0.059 & 0.049 & ... & ...  & ...  & 0.292 & 0.158\\
\hline
0.3 & 19 & 0.98  & 7.715 & 0.28 & 0.183 & 15.4 & 0.08 & 0.051 & ... & ...  & ... & 0.294 & 0.191 \\
\hline
0.5 & 22  & 0.92 & 8.2 & 0.108 & 0.021 & 16.41 & 0.016 & 0.009 & 19.53& 0.016  & 0.014  & 0.212 & 0.112 \\
\hline
0.7  & 25 & 0.87  & ... & ...  & ...  & ...  & ... & ... & 20.31 & 0.099 & 0.03 & 0.102 & 0.032\\
\hline
1  & 28 & 0.81 &...  & ... &  ...& ... & ... & ... & 21.58 & 0.08 & 0.028 & 0.082 & 0.030 \\
\hline
1.2 & 30 & 0.77 & ... & ... & ... &  ...& ... & ... & 22.36 & 0.062 & 0.024  & 0.064 & 0.026\\
\hline
1.5 & 33 & 0.73  & ... & ... & ... & ... & ... & ... &  23.44 & 0.004 & 0.002 & 0.004 & 0.002 \\
\hline
\end{tabular}
\label{table_mdot10}
\end{table}
\end{onecolumn}

\end{document}